HACK NDSU: A Real-world Event to Promote Student Interest in Cybersecurity

Enrique Garcia
Information Technology Division and Department of Computer Science
North Dakota State University
1320 Albrecht Blvd., Room 258
Fargo, ND 58102
Phone: 701-231-5590
Email: enrique.garcia@ndsu.edu

Jeremy Straub[1]
University of West Florida
220 W. Garden St., Suite 250
Pensacola, FL 32502
Phone: 850.474.2999
Email: jstraub@uwf.edu

**Abstract**

Hack NDSU let students scan, probe, and hack North Dakota State University's campus network, under professionals' supervision, providing an aspirational experience, potentially motivating them to enter the field. This paper provides a blueprint for educational hacking events against production systems. No prior educational event of this type is known.

---

[1] This work was conducted while J. Straub was at North Dakota State University.

# 1. Introduction

Digital transformation has increased society's reliance on computer systems. Nowadays, it is imperative to consider cybersecurity as an essential component of every process in IT. It is not uncommon to hear news reports about hacking and breaches, as more operations and daily activities rely on computer systems and electronic devices. The attack surface is increasing, as more everyday devices are connected to the internet.

Internet connected systems bring many benefits, but they also create challenges by exposing the systems to attacks from any place in the world where a bad actor has online access. A shortage of skilled cybersecurity professionals has developed. According to (ISC)², there is a global unfulfilled demand of 4.7 million cybersecurity professionals, with about 550,000 of these jobs located in North America [1]. The shortage of cybersecurity professionals is only expected to get worse. (ISC)² has estimated that, in order to meet the global demand for cybersecurity professionals, the existing pool of 5.4 million workers needs to grow by 87% [1].

ISACA's 2024 "State of Cybersecurity" study said that 27% of security professionals surveyed indicated that university graduates are qualified and 73% of respondents said that prior hands-on experience is the most important qualification for a cybersecurity candidate [2]. Educational institutions' new graduates are still the main source of recruitment for cybersecurity professionals [3] which means universities have a responsibility to close the gap, both in terms of production quantity and by choosing teaching techniques and tools that are well suited for cybersecurity education. The U.S. Department of Commerce "Report to the President on Supporting the Growth and Sustainment of the Nation's Cybersecurity Workforce: Building the Foundation for a More Secure American Future" recommended that cybersecurity-related education programs build on and strengthen "hands-on, experiential and work-based learning approaches—including apprenticeships, research experiences, co-op programs, and internships" [4]. A study conducted by the Center for Strategic and International Studies (CSIS) surveyed IT professionals from eight countries and the respondents said that "practical training and hands-on experience is necessary to equip students with the tangible skills employers expect" [5]. The IT professionals also noted that "incorporating practical learning into academic programs would better prepare cybersecurity professionals for the real world" [5].

The message from the IT field is clear: There is a shortage of IT security professionals and college students require hands-on skills for the security challenges that the industry faces.

# 2. Hack NDSU

This section provides an overview of Hack NDSU. Section 2.1 introduces Hack NDSU, and Section 2.2 explains the significance of the event. Finally, Section 2.3 describes the educational methods used.

## *2.1. The Hack NDSU concept*

Hack NDSU started as a collaboration between the NDSU Computer Science Department and the NDSU IT Security Office during the 2019 spring semester. The premise was to invite professional penetration testers to campus to demonstrate and teach students how to scan the network, probe for vulnerabilities, and hack the NDSU IT systems. The NDSU Vice President for IT (VPIT) was very enthusiastic from the moment the idea was pitched to him. The VPIT saw the value for the students, in the form of a practical learning experience, and for the IT Division in the form of a security assessment with a very low cost to the university.

From the academic perspective, the goals of the exercise were to get the students excited about cybersecurity, show them the adversarial nature of the field, have them gain real world experience in IT security, and learn the process and tools used for penetration testing. In addition to acquiring technical skills, students would learn the administrative, legal, and governance aspects of a project of this type. The 2019 edition of Hack NDSU helped identify the dynamics of the event and provided insights allowing adjustments, in order to use the 2020 edition for research purposes.

*2.2. Significance of the research project*

The event took an innovative approach. Other institutions allowing students to hack their production systems are not known to have occurred. Hack NDSU provided real world, hands-on, practical experience with live systems, unlike cyber range simulations. Additionally, Hack NDSU taught cybersecurity from the offensive – rather than defensive – perspective. Krank [6], a member of the armed forces, argued that learning cybersecurity from a defensive perspective does not stimulate the adversarial mindset needed to become a cybersecurity professional. Finally, most organizations are hesitant to allow students or individuals without experience to scan and probe their network and systems for vulnerabilities. Because of that, the event was carefully planned and strict rules of engagement were defined to ensure the stability of the systems, while still allowing the students to find, exploit and document vulnerabilities without compromising the systems' availability and integrity.

This event required a combination of logistics, inter-departmental coordination, agreements, developing content, and addressing legal requirements, such as having the students sign non-disclosure agreements and obtaining written approval from the NDSU VPIT. The NDSU IT Division also signed a penetration testing agreement with the local company that donated professional services to facilitate the activity. Those agreements had to be reviewed by legal counsel and the Office of the Attorney General.

The NDSU IT Division has an on-call rotation to respond to alerts and alarms. They were informed and consulted about the event and agreed to provide support to address any possible service disruption.

The NDSU Computer Science Department and the NDSU Institute for Cyber Security Education and Research provided assistance with logistics, promotion, equipment, and guidance.

The event concept can have a wide range of applications. It can be the model for a series of events that would address areas of cybersecurity such as physical security or web application security. With the encouraging initial results, this type of event can be used by other educational institutions to generate interest on cybersecurity which could contribute to closing the workforce gap.

The event also addressed feedback from the IT industry. Surveys [5] show that the industry believes real world activities are needed to better educate cybersecurity students. The event built on that feedback and provided students with a resume builder, showing that they have hands-on experience in a real world production environment.

In summary, Hack NDSU provided students with the opportunity to participate in an offensive, real world penetration testing engagement with unscripted scenarios in a legally safe environment under the direction of knowledgeable security professionals.

*2.3. Traditional and nontraditional learning methods*

Evidence shows that a blend of lectures and active learning can provide favorable results [7]. Thus, Hack NDSU mixed interactive lecturing and several active learning methodologies. It had a heavy emphasis on a problem based learning, laboratory and gamification approaches.

The event started with a lecture session to provide background information that students hadn't covered in their classes. The three-hour lecture provided a foundation and mirrored professional boot camps. The lecture covered the legal aspects of ethical hacking, numeral systems, networking, network scanning, and cryptography. The orientation session used a PowerPoint presentation, the Socratic method, and active learning activities designed to keep the students engaged. Some of the didactic materials used were envelopes to demonstrate encapsulation on the Open Systems Interconnection model, stress balls to demonstrate the Transfer Control Protocol/Internet Protocol handshake, and tent cards to simulate hosts in a network. Participation was rewarded with small vendor prizes. The prizes, food, beverages, and the freedom to move around the classroom fostered a low risk environment that encouraged student participation.

The hacking event was performed against live systems with the involvement of IT security professionals. It reinforced the concept of situational learning by providing a genuine, real world environment [8]. It also showcased the culture of one branch of cybersecurity, including the jargon and other aspects that may seem insignificant – such as the stickers the pen testers had on their laptop lids, their attire with corporate logos, the vendor giveaways, and the anonymized field stories the professionals shared.

The event embraced the concept that "learning is not a spectator sport" [9]. Hack NDSU aimed to tie existing knowledge to the event experience to enhance learning, retention of the concepts, and transferability to other scenarios, as suggested by [10].

### 3. Logistics

This section discusses the logistics of the event. Section 3.1 discusses the non-disclosure agreements signed by the students. Section 3.2 covers the agreement with the local IT company that assisted during the event. Finally, Section 3.3 discusses the resources needed during the event.

#### *3.1. Student non-disclosure agreement*

A non-disclosure agreement was signed by the students participating in the event. The agreement had the dual purpose of demonstrating part of the legalities of a penetration testing engagement and protecting sensitive data that the students could have access to during the event. The students were provided with IP addresses and ranges for the NDSU network, including mission critical equipment. Since it was a test against production systems, there was the potential that the students could run into usernames and passwords, user files, and other sensitive data.

The 2020 version of the non-disclosure agreement was changed to include a statement to explicitly allow students to advertise their participation in the event, so they could use it as a resume builder. It allowed the students to list the event on their resume and talk about it to recruiters in general terms but prohibited disclosing particular details.

Another change from the first edition of Hack NDSU was changing the non-disclosure agreement to allow students to take notes and write down information about tools and other general knowledge. The 2020 non-disclosure agreement included providing the students with paper of two different colors for note taking. Plain white paper was used for taking general notes that did not contain sensitive information that the students could take it home after it was reviewed by the organizers. Green color paper was used to write down sensitive information such as IP addresses, network ranges, devices, usernames and passwords. The green paper was collected at the end of each night and deposited into a secure document disposal bin.

Students signed the agreements as they came into the room for the event. The organizers sent the agreements to the attorney to sign and the students were provided a counter-signed copy.

#### *3.2. Penetration Testing Agreements*

The NDSU IT Division and the Computer Science Department signed agreements with a local IT firm with cybersecurity staff' that provided penetration testing services.

The first agreement was a master services agreement which spelled out the obligations of each organization and included liability, indemnification and confidentiality clauses. The master services agreement was a general service agreement with no penetration testing specifics.

The second agreement was an ethical hacking agreement which defined the scope of the event and added clauses limiting liability. The agreement also stated that the pen testers would not exploit any vulnerability that would lead to a denial of service, unless it was requested in writing by NDSU. Additionally, the agreement contained a clause allowing the local IT company's employees to use their findings from the event to educate other organizations about lessons learned without revealing where any vulnerabilities were discovered. This clause is similar to the clause added to the student non-disclosure agreement to allow resume builder use.

#### *3.3. Resources needed*

Many hours were spent planning and coordinating the event. In the first edition of Hack NDSU, a considerable amount of time was spent on logistics, and coordinating meetings with the NDSU IT Division and the pen testers. With the lessons learned from the first iteration of Hack NDSU, adjustments were made, requiring a fair amount of time to implement.

During the event, the students used laptops and Raspberry Pis from the Computer Science Department. Time was spent creating materials for the interactive lecture and hacking/vulnerability exploitation sheets. A classroom with multiple screens for the presenter and power strips were also used. The laptops and Raspberry Pis connected using the wireless network, so there was no need for wiring a network. The small prizes and giveaways were provided by vendors and collected from conferences by employees of the IT Division who donated them for the event. The only financial cost was for the food and beverages for the student participants.

The local IT company donated their time and pen testing expertise. The knowledge and experience of the pen tester was a great asset for each event and more than one student commented on the passion of the pen tester for his field.

The costs to the IT Division came in the form of time spent by the employees in meetings, some technical tasks such as safe listing IPs, and ensuring the on-call staff was aware of the event. NDSU has a hybrid IT model where central IT provides many common services and some departments hire IT professionals who work directly for them. Those departments benefited from the event indirectly, since sometimes new devices are connected to the network and not properly secured. By finding those devices, the project assisted those departments by identifying the vulnerable devices and notifying the system owner.

**4. Penetration Testing**

This section discusses penetration in the context of Hack NDSU. First, Section 4.1 discusses a common methodology. Then, Section 4.2 discusses the methodology used for Hack NDSU.

*4.1. Common penetration testing methodology*

There are many types of penetration tests for areas including social engineering security, physical security, wireless security, telecommunications security, and networking security. Most of the methodologies adhere to the following general steps [11]:

In the reconnaissance stage, the attackers try to learn as much as possible about their target including IP ranges. This can be done by visiting the target's or third-party websites, identifying their customers, determining their email addresses format, and looking at their job ads.

In the scanning stage, the attackers use the IP ranges found in the first step and identify which computers respond to requests. Detection can prospectively be avoided by slowing the scanning rate below typical alerting system thresholds.

In the fingerprinting stage, attackers try to find details about hosts' services, such as the software used and its version.

The next step, host analysis and vulnerability research, determines whether there are vulnerabilities that can be exploited on a host. This includes researching default usernames and passwords for the system, as leaving default credentials unchanged is a common mistake. Social engineering, which requires little technical knowledge, can also be used.

In the exploiting the vulnerability and gaining access step, the information obtained in the prior step is used to obtain access to the system.

Once the attacker has gained access to the system, the next step is to ensure continued access. Some hackers patch systems, to avoid other hackers gaining access, change default credentials, or create secondary administrator accounts.

In the final attack step, covering tracks, the attacker seeks to mask their attack by deleting system logs, either in entirety or selectively and through other similar techniques.

Finally, reporting is typically required for penetration testing. The report normally includes findings, the severity of the vulnerabilities found, and mitigation recommendations.

*4.2. Penetration testing methodology used during Hack NDSU*

Hack NDSU followed most of the common steps discussed in Section 4.1. Figure 1 shows the process.

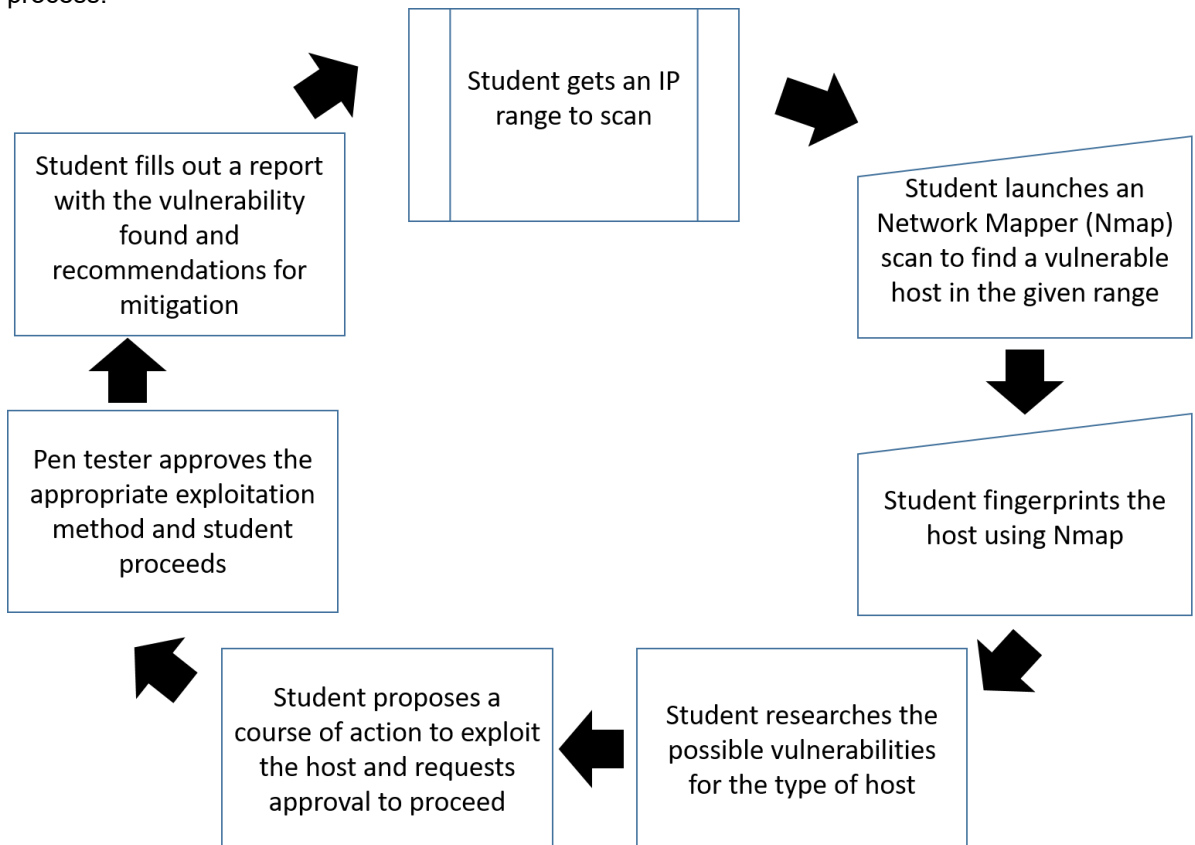

Figure 1. Hack NDSU penetration testing methodology.

First, to expedite the event, the reconnaissance stage was skipped, as the target network and network ranges were known.

Scanning was next. The organizers provided the network ranges to the pen testers and scanning started two weeks prior to the event. The pen testers conducted both internal and external network scans. The NDSU IT Division safe-listed the IP used by the pen testers to avoid alerts or blocking his scanning device. The local IT firm provided a computer for the internal scan to connect to the NDSU network with commercial network scanning software that provides a vast range of information about vulnerabilities it finds on hosts along with a score of the level of risk.

Scanning is time consuming. During the first edition of Hack NDSU, the pen testers had to fill time by telling stories during the initial network scans. While the stories were entertaining, students can become disengaged if they are idle for too long. In the second edition, this step was performed in advance to create a more dynamic environment. By starting two weeks in advance, the pen testers were able to scan the whole NDSU IP network range, rather than just parts of it. To give the students a realistic penetration testing experience, the students started with scanning and ended with exploitation and reporting. The exercise used an IP address range with a known-vulnerable host. The students scanned this range using Network Mapper (Nmap) and found the hosts in a few minutes. This scaled version of a real scan provided a satisfactory experience in a reduced amount of time. The organizers worked with the NDSU IT network enterprise team to safe list the IP addresses used by the students to avoid them getting blocked in the middle of scanning.

Fingerprinting was performed third. One of Nmap's features is the script engine that has many task automation scripts. To expose the students to this tool, once the students located the host, they used the "banner" script to determine which services the host was running, based on the open ports. This also identified the software possibly running on the device.

Next, analysis and vulnerability research was conducted. The output from the "banner" script provided a fairly accurate assertion of what the host was. This gave the students direction to research vulnerability databases for possible attack channels. For web servers, they also browsed to the IP address and collected additional information to search for the default username and password.

Exploiting the vulnerability and gaining access is the fifth step. The agreed rules of engagement were that the students would provide proof of concept but would not take any action that would adversely affect NDSU. Once the students determined a course of action, they proposed the approach and, if it was adequate, the students performed the exploit to obtain proof of concept.

Most students only attempted to log in to the hosts. Some found default credentials online and other guessed the password. Every time a student was successful, they received public recognition and a small prize. The achievement was celebrated in front of all the attendees and the student was asked about the details of the exploit. The rest of the room cheered and then the student started over again with a different network range. Students said that the public praise was very well received and was a subtle way of turning the event into a friendly competition.

The maintaining access step was not performed. IT contacts were made aware of issues to patch or secure the devices with identified vulnerabilities.

Similar to the previous stage, the covering tracks state was not performed.

Reporting was the final task. The students wrote a brief report on the vulnerability they found and recommendations to address or mitigate the problem. Anecdotal information from pen testers highlighted the importance of the skill while also indicating it is the most boring and dreaded part of penetration testing.

**5. Discussion**

This section evaluates and explores the benefits of the Hack NDSU event. Section 5.1 discusses the key take-aways from the event. Section 5.2 explains its contributions to the cybersecurity field. Section 5.3 presents the lessons learned during the execution of the event. Finally, section 5.4 discusses the limitations of Hack NDSU.

*5.1. Event Take-Aways*

Student feedback was clear that hands-on activities are preferred over passive activities. They expressed that they enjoyed the hands-on activities and that they preferred walkthroughs rather than demonstrations. Resnick uses the term playful learning to describe the cognitive process that occurs when an individual engages in activities that are enjoyable and stimulate learning as a consequence [12]. Hack NDSU had many elements of a game. Students had a chance to hack a device and be rewarded with public recognition and a small prize. They acquired cybersecurity knowledge, personal satisfaction, experience penetration testing live systems, and a resume builder.

The live environment gave validity to the event and made it attractive. Several students perceived Hack NDSU as a non-academic activity, similar to ethical hacking in the real world. The live environment also provided immediate feedback which is conducive to learning. The professional penetration tester was an invaluable asset. The students knew they were working with a person with real experience and the individuals had plenty of stories to share. A student commented that "having an opportunity to practice on a live environment…helped the students gain confidence on their abilities".

Hacking systems that the students were familiar with made an impression on them. A student commented that they will never see a certain facility the same way after knowing that they hacked a device hosted in it. Another student was surprised to learn that devices used in production were vulnerable and how some of the weaknesses were something as simple as default username and passwords. Yet another student mentioned that employers often criticize academia for not providing students with practical skills and that the event provided an opportunity to gain real world skills for their resume.

The event also highlighted the need to remove as many barriers to learning as possible. Several students were frustrated with the performance of the Raspberry Pis and commented how underpowered they were. The University of Maryland Baltimore County held an exercise where students examined a copy of a portion of a production network and received similar feedback regarding power strip issues [13]. Small setbacks, such a slow or unresponsive Raspberry Pi or having to difficulty finding a power outlet can lead to ego depletion which can impair learning and task performance [14]. Another student mentioned that the uncertainty of the Coronavirus pandemic was in their mind during the event. A survey showed that the pandemic negatively affected 50% of students' mental health and ability to learn [15]. There are some circumstances that cannot be controlled, so it is advisable that organizers try to remove as many obstacles as possible.

Students need a foundation for the event to be enjoyable and successful. Participants didn't prepare for Hack NDSU, underscoring the importance of the orientation session providing a limited background. Students made comments such as they would have been lost without the orientation session and that the knowledge acquired in the session made days two and three possible. The pre-event evaluation surveys show that students lacked the knowledge necessary to understand the penetration demonstration and to hack on their own. A lot of work and planning was required to make it an active learning session, full of activities, to keep the students engaged. The feedback on the orientation session was positive and showed the importance of having it.

*5.2. Contribution to the cybersecurity field*

As far as is known, Hack NDSU was the first academic event where students scanned, probed, and hacked a production university network with the assistance of penetration testing professionals. This provided students with practical experience that industry and recruiters desire from college graduates.

Hack NDSU showed that this approach can help educators spark interest in cybersecurity, debunk misinformation about the field, and hone students' adversarial mindset. In combination with curriculum and other events, this kind of activity can help attract students to the field and help close the cybersecurity labor gap.

There is also the potential to use the Hack NDSU model in the private sector. No incidents or service disruption occurred, as the students followed the directions and rules of engagement laid out to them. Students were closely monitored by the professional penetration tester and the organizers. This controlled approached helped the students understand cybersecurity ethics. Even though the exercise ran against a school network, it is still a production IT network and not a copy or a simulation. The NDSU network runs mission critical services, similar to private sector networks. Informal conversations with Chief Information Officers (CIOs) from the private sector, provide anecdotal data that CIO's would be hesitant to allow students to learn using their production systems. Hack NDSU showed that this type of event can be conducted in a controlled and safe manner. It can potentially help companies train the future workforce and provide access to the pool of future graduates. Both students and organizations would benefit from similar exercises: students would get practical experience and companies would get a free or low-cost security assessment. Similarly, a company could contract with a professional pen tester and conduct an exercise similar to Hack NDSU with their employees attending for training purposes.

*5.3. Lessons learned*

Students were free to hack into any system any way that the pen tester or organizers approved. More complex hacks, however, would pose a challenge. For instance, a SQL injection or cross site scripting attack would have required those topics to be included in the orientation session, requiring more time. The purpose of Hack NDSU was to provide hands-on activities, no extensive classroom instruction, making advanced attacks impractical unless students have prior preparation

The production environment created uncertainty regarding which kinds of vulnerabilities will be found. Unlike a prepared scenario, there is no certainty on whether, how many, or which type of vulnerable hosts the students will find, so it is not feasible to prepare the students for every scenario. This is where the professionals involvement is important and collaboration with the IT department is crucial. Students also learn to deal with unexpected circumstances.

Hack NDSU required a large student time commitment. It took place over three nights and lasted 12 hours. Most, but not all, students attended all three nights. While this was a big time commitment, students expressed that it was worth it.

### *5.4. Limitations of the research project*

Participants started with different levels of knowledge. All students had a basic working knowledge of computer systems and the orientation session built on that knowledge. While all students hacked into at least two systems and all expressed enjoying the event, their level of knowledge and preconceived ideas of cybersecurity may have affected how they perceived the event.

It was difficult to find lots of students to participate in the event (and only a subset agreed to an interview). It would be ideal to have a follow up event and interview participants a year after the event.

## 6. Conclusions and Future Work

Hack NDSU proved to be a valuable exercise for the students and the cybersecurity field. The event generated positive perceptions towards cybersecurity and most participants expressed enjoyment and excitement towards the field.

Students learned that a basic understanding of IT and a night's worth of training can allow basic network attacks to be conducted. This expanded the students' perception of cybersecurity, both professionally and personally. Students realized that many systems lack proper cybersecurity. Some realized that the way they broke into the vulnerable systems could be used to compromise their own personal devices and accounts, making them think about their own personal cyber hygiene. Additionally, some students expressed plans to run penetration testing against their own devices to find out if they had vulnerabilities.

Understanding the state of cybersecurity and need for personal cyber hygiene are important realizations, given our dependence on information systems.

Hack NDSU also helped students see past the unrealistic picture the entertainment industry has painted of hacking. They got a realistic perspective of what it is like to be a cybersecurity professional and the skills needed. The event helped the students understand hacking as something they can explain to others and do themselves. Several students expressed having a better understanding of hacking and feeling better informed about cybersecurity. By having a more realistic perception, students can make better career decisions and realize that they can become cybersecurity professionals, with the right training.

Students also became more confident in themselves and their skills. During Hack NDSU they did what they had previously only heard of or read about. They learned that with the right kind of training they could hack a live network. Overall, the event was successful and valuable to the student participants. In the future, it could be expanded more broadly within North Dakota and beyond and into the private sector worldwide.